# Towards Digital Twins for Optimal Radioembolization

Nisanth Kumar Panneerselvam, PhD[1*], Guneet Mummaneni, M.S.[1*], Emilie Roncali, PhD[1,2]

[1] Department of Biomedical Engineering, University of California Davis, Davis, CA USA

[2] Department of Radiology, University of California Davis Health, Davis, CA USA

[*] Equal contributors

Corresponding Author: Emilie Roncali

Email: eroncali@ucdavis.edu

## Key words

## Key points

- Radioembolization is increasingly using targeted high radiation doses, which requires precise treatment planning based on dosimetry
- Treatment planning consists in selecting the best injection points and injected activities, which can be optimized through modeling the microsphere transport in blood and dose distribution
- Computational fluid dynamics (CFD) can model the microsphere blood transport and distribution precisely to form the basis of a liver digital twin for radioembolization

- CFD becomes computationally prohibitive if repeated simulations are necessary to sample multiple injection point locations within the liver arterial tree
- Generative AI and physics-informed neural networks (PINNs) in particular offer a promising alternative to CFD simulations to rapidly generate realistic microsphere distributions while maintaining simulation accuracy

# Synopsis

Radioembolization is a liver cancer treatment delivering radioactive microspheres (20~60 µm) to tumors via a catheter in the hepatic arterial tree. Treatment response depends on multiple factors including the complex hepatic artery anatomy, variable blood flow, and microsphere transport, which should be considered in treatment planning.

Patient-specific digital twins powered by computational fluid dynamics (CFD) and physics-informed AI methods offer a promising solution to optimize planning.

This review discusses core principles of CFD and generative AI applied to radioembolization, emphasizing physics-informed networks and their role in translating digital twins into clinical practice for enhanced personalization and precision in treatment delivery.

# 1 Background:

## 1.1 Transarterial embolization for liver cancer

Liver cancer remains one of three cancers with rising incidence and mortality [1]. Aside from primary tumors such as hepatocellular carcinoma (HCC), the liver is a key metastatic site for colorectal, breast, pancreatic, and neuroendocrine cancers [2]. Transarterial embolization uses microspheres, particles, or emulsions injected in the liver bloodstream through a catheter to obstruct tumor blood flow, deliver localized radiation (radioembolization) or chemotherapy (chemoembolization) [3], [4], [5]. Performed in interventional radiology over at least two sessions each taking a few hours, transarterial radioembolization, also called Selective internal radiation therapy (SIRT), includes a workup followed by one or more treatment sessions 1-2 weeks apart. The workup, useful to build the treatment plan, consists of vascular mapping with cone-beam CT (CBCT) and estimation of lung shunting [6].

The efficacy of transarterial embolization depends on vascular anatomy, liver and tumor blood flow, and venous shunting. Liver tumors primarily receive blood from the hepatic artery, unlike the rest of the liver that gets most of its blood supply from the portal vein—allowing transarterial embolization to selectively target tumors [7]. Radioembolization utilizes 20-60 µm microspheres made of glass or resin containing radioactive yttrium-90

($^{90}$Y); its efficacy is best assessed by the absorbed radiation dose. However, tumor targeting is highly patient-dependent and difficult to plan and implement with current imaging techniques. Personalized, precise planning could significantly improve transarterial embolization's impact on patient outcomes, making treatment simultaneously safer and more effective [8], [9].

## 1.2  Importance of treatment planning and challenges

Radioembolization planning relies on injection site, microsphere quantity, vascular anatomy, and tumor blood flow, which collectively determine the microsphere and resulting radiation dose distribution. Standard-of-care imaging, including contrast-enhanced CT (CECT) and CBCT provides vascular and blood flow information with a resolution of 200-500 µm, defined by the CT voxel size and the contrast to noise ratio in arterioles. The tumor vascularization thus directly affects the size of vessels that can be segmented. $^{99m}$Tc macroaggregated albumin SPECT can be used to predict the $^{90}$Y microsphere dosimetry with good accuracy in the healthy liver but with errors exceeding 100 Gy in tumors in 5% of patients [10], [11]. While these errors are primarily due to differences of injection locations, they may also be caused by differences between $^{90}$Y microspheres and biodegradable $^{99m}$Tc-MAA (irregular particle shape, size distribution, density). As radiation segmentectomy becomes predominant with extremely high doses (>190 Gy) to small volumes, precise and personalized planning is increasingly critical [12], [13].

## 1.3  Modeling approaches and digital twins

As the treatment response depends on the number of microspheres reaching the tumors, predicting their distribution in the liver is the cornerstone of robust and precise treatment planning. Predicting the microsphere distribution before treatment presents several unresolved challenges. Due to their size and density, microspheres can be assumed to follow the blood flow only in large arterial branches, making measuring or modeling blood flow only insufficient to evaluate their distribution in small distal vessels and likely requiring computational models with higher fidelity. In addition, placement of the catheter is critical as the distribution of microspheres is extremely sensitive to it (Figure 1, [14]). Multiphysics modeling combining computational fluid dynamics (CFD) and radiation physics has been investigated by several groups as an alternative to standard-of-care treatment planning [15], [16], [17], [18], [19].

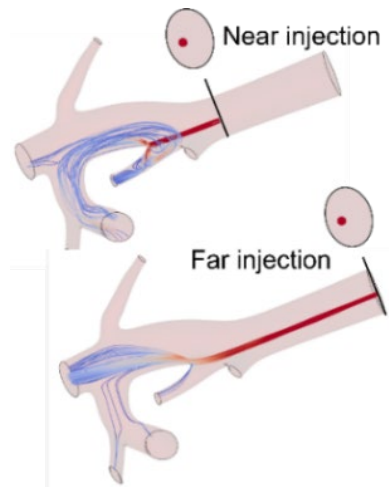

**Figure 1. The injection location strongly directs the blood flow between branches and should be carefully planned.**

Such transport models, informed with images and data, can be personalized for each patient and integrated into a theranostics liver digital twin. The patient's liver twin can be operated to optimize the injection location and activity to obtain the desired radiation dose distribution. Digital twins have been employed in various engineering problems ranging from designing manufacturing production lines, hospital workflows, and aerospace [20], [21], [22]. A digital twin is an operator described by mathematical or computational models (e.g. partial differential equations, Monte Carlo simulations, computational fluid dynamics) that are personalized via specific input data. In healthcare, digital twins are biomedical models informed with patient data. These data are not limited to images and may include various biomarkers indicative of liver function, specific tumor composition information, or vascular markers. The goal of the digital twin is to predict the behavior of the patient or organ— in our case the liver — in response to stimuli, such as the injection of radioactive microspheres in radioembolization. Its output is not limited to the injection point and activity for treatment planning and could expand to tissue response to radiation or disease progression prediction [23], [24], [25], [26], [27], [28], [29].

Building and validating digital twins presents challenges in data standardization between institutions, computational infrastructure, model accuracy, and patient-specific implementation. In radioembolization, digital twins can employ CFD, using a 3D mesh of the vasculature extracted from patient images to model the blood and microsphere transport

(see section 2 for CFD principles). This poses a technical challenge, as high-fidelity CFD rapidly becomes computationally prohibitive, especially in a large geometrical structure such as the liver arterial tree requiring millions of mesh elements. AI-based models may be able to accelerate computation, creating exciting research opportunities. Specifically, physics-informed neural networks (PINNs) can alleviate the data requirements by leveraging physics instead of requiring large training sets [30], [31], [32].

In this review, we discuss three aspects of liver digital twins: CFD principles and challenges; AI solutions to accelerate simulations; combining AI and physics to achieve optimal results with PINNs.

## 2  Computational methods

### 2.1  Computational fluid dynamics

The four main phases of CFD are 1) geometry modeling, 2) discretization, and 3) boundary conditions 4) governing equation solving. Each phase is described below.

#### 2.1.1  Geometry: Imaging to patient-specific 3D model generation

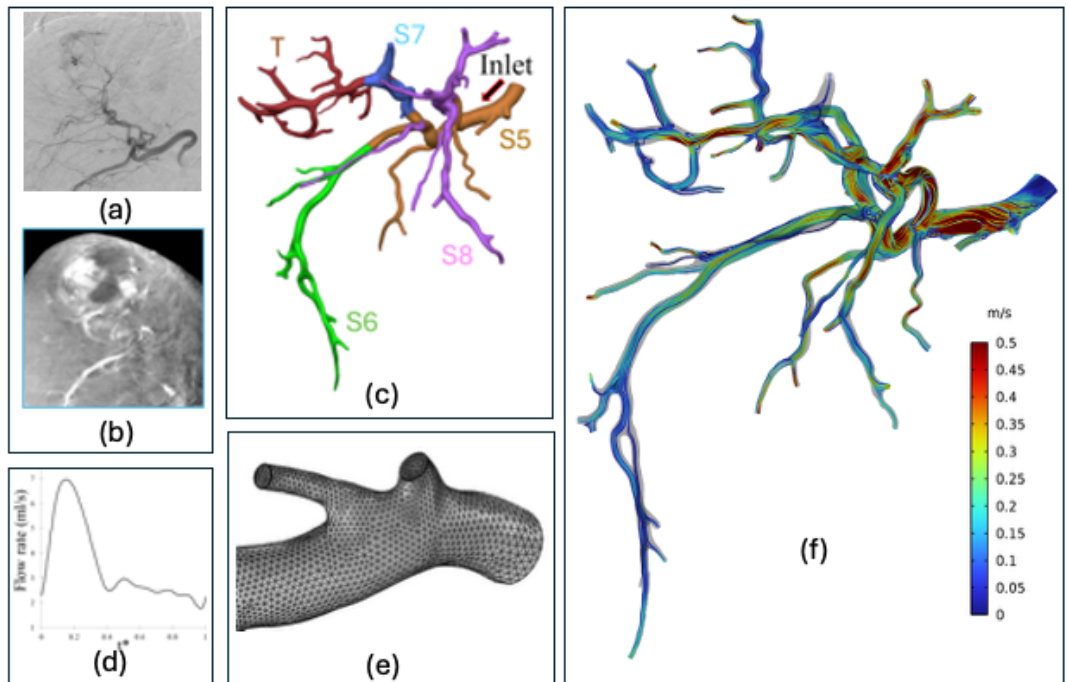

Figure 2. (a, b) CBCT and CECT images used for hepatic artery reconstruction. (c) Hepatic artery 3D model used as flow domain. (d) Pulsatile wave of inlet flow rate (ml/s). (e) Corresponding mesh of flow domain. (f) Blood flow in hepatic artery tree represented by streamlines.

The hepatic artery network in humans is highly complex and tortuous; each individual has a unique hepatic artery tree structure [33]. Although generic anatomical models serve a purpose in fundamental research, they fail to reflect the subtle hemodynamic variations that control microsphere transport in specific patients. Alternatively, a patient-specific 3D model of hepatic artery trees can be generated from high-resolution medical imaging [34], typically from CT or cone-beam CT (CBCT) as shown in Figure 2 (a) and (b). In this figure, Roncali et al. [15] developed the hepatic artery tree 3D geometric model (Figure 2 (c)) from CBCT images using a custom marching cubes algorithm and the open-source Vascular Modeling Toolkit (VMTK). Reconstruction of the geometry model from medical images involves image segmentation, surface smoothing, and simplification of inlets and outlets of vasculature. These are crucial geometry reconstruction steps, which directly influence the accuracy of CFD simulation outcomes. Recent developments in deep learning have facilitated its broader application in medical image analysis, which has led to substantial enhancements in segmentation accuracy [35]. Some of the main AI-based image segmentation techniques are detailed below:

- U-Net is a convolutional neural network (CNN) framework specifically engineered for biomedical image segmentation applications, presenting the advantage of relying on limited training datasets [36].

- V-Net is a fully convolutional 3D architecture tailored for segmenting volumetric medical imaging data [37].

- nnU-Net is a self-adaptive deep learning framework that automatically optimizes preprocessing, network design, training, and postprocessing steps for diverse biomedical segmentation tasks [38].

- Attention U-Net enhances the U-Net architecture by incorporating attention gates, which enable the model to emphasize clinically relevant spatial regions such as small vessels or lesions while reducing the influence of irrelevant background features [39].

- DeepLabv3+ integrates atrous convolution with spatial pyramid pooling to enhance multi-scale feature representation to improve complex anatomical structures delineation [40].

2.1.2   Discretization: Hepatic artery 3D model meshing

After segmenting the 3D image to generate the patient-specific 3D fluid domain model, the next step involves transforming this model into a 3D volumetric mesh accurately reflecting the patient-specific flow domain's shape and size, featuring a smooth surface mesh (Figure 2 (e)). Meshing represents the discretization phase where the continuous and complex mathematical equations governing fluid flow are converted into a set of discrete, computationally solvable equations. The meshing process divides the flow domain into an extensive array of smaller, interconnected elements forming a computational mesh or grid. Each element in this mesh corresponds to a discrete volume where the fluid dynamics equations are numerically solved. Various techniques for creating the initial mesh for CFD biomedical application are outlined in [41].

### 2.1.3 Patient–specific boundary conditions

The incorporation of patient-specific boundary conditions is crucial in radioembolization CFD simulation for analyzing personalized intrahepatic hemodynamics [19], [42]. In vascular CFD models, boundary conditions specify the blood and vessel environment interactions at the domain edges, influencing flow and pressure distributions. The clinical relevance of simulation results is highly dependent upon patient-specific boundary conditions. Due to the compliance and resistance of the hepatic artery, the boundary conditions at the upstream and downstream ends of the arterial tree are entirely unknown. Several studies have employed patient-specific hepatic artery flow rates (shown in Figure 2 (d)) measured by transcutaneous Doppler sonography, a method recognized for its higher accuracy over generalized waveforms [43], [44]. The simplest form of defining outlet boundary conditions is zero pressure at outlets; this facilitates traction-free exit of fluid flow through multiple branches. The 3-element Windkessel model is a common method for calculating patient-specific outlet boundary conditions, which includes the influence of downstream vasculature [45].

### 2.1.4 Governing equations and solvers for radioembolization

In CFD, a set of fundamental governing equations are used to model the hemodynamics of blood flow and the transport of $^{90}Y$ microspheres within the hepatic arterial system. The primary equations begin with the continuity equation (conservation of mass), ensuring mass conservation throughout the vascular domain:

$$\nabla \cdot \vec{u} = 0 \quad (1)$$

where $\vec{u}$ is the velocity vector of fluid, this ensures zero net accumulation of mass inside the control volume. This is followed by the Navier-Stokes equation, ensuring momentum conservation based on Newton's second law in fluid dynamics.

$$\rho \left( \frac{\partial \vec{u}}{\partial t} + (\vec{u} \cdot \nabla)\vec{u} \right) = -\nabla p + \mu \nabla^2 \vec{u} + \vec{F} \quad (2)$$

where $\rho$ is fluid density, $\mu$ is dynamics viscosity, $p$ is pressure and $\vec{F}$ is the body forces.

In radioembolization the diameter of the glass and resin microspheres is 20-30 μm and 20-60 μm, respectively. The specific activity per microsphere is higher in glass microspheres (2500 Bq) than resin (50 Bq). Variations in diameter and specific activity influence the embolic effect and radiation distribution of microspheres [46]. The prediction of $^{90}$Y microsphere transport is based on the fundamental governing equations behind blood flow simulation. The resulting blood flow field computed by the governing equations serves as the foundation for particle tracing simulations (shown in Figure 2 (f)). The one-way and two-way coupling methods define the interaction between blood and microsphere while performing particle tracing simulation of microsphere distribution in radioembolization.

In the one-way coupling approach, the incompressible Navier-Stokes equations (Eq. 2) are first solved to establish the fluid flow field. Subsequently, microsphere trajectories are computed based on a pre-determined fluid field with the key assumption that the microspheres do not influence the fluid flow. Following the determination of the flow field, particle motion is described using Newton's Second Law formulated in a Lagrangian framework [47]:

$$m_p \frac{d\vec{v}_p}{dt} = \vec{F}_{\text{drag}} + m_p \vec{g} \qquad (3)$$

where $m_p$ is mass of microspheres, $\vec{v}_p$ is velocity of microspheres, $\vec{F}_{\text{drag}}$ is drag force, and $\vec{g}$ is gravitational acceleration vector. One-way coupling simplification is preferable for significantly low concentrations of dispersed particles, which have a negligible impact on the bulk fluid [48]. In the two-way coupling model, the fluid influences microspheres, and microspheres affect the fluid flow through bidirectional momentum exchange between the fluid and microspheres. Therefore, the source term that represents the momentum transfer from particle to fluid is included in the modified Navier-Stokes equation [49].

$$\rho \left( \frac{\partial \vec{u}}{\partial t} + (\vec{u} \cdot \nabla)\vec{u} \right) = -\nabla p + \mu \nabla^2 \vec{u} + \vec{F} + S_p \qquad (4)$$

where $S_p$ represents the force exerted by microspheres on the fluid. The two-way coupling model is a near-realistic approach preferred in digital twins of radioembolization, where the concentration of microspheres affects the blood flow locally. The four-way coupling model, which includes microsphere-microsphere interaction, increases computational complexity. Incorporating a more comprehensive representation of microsphere dynamics, specifically microsphere-to-microsphere interactions, into computational models is anticipated to yield more accurate predictions of microsphere accumulation. This enhanced microsphere accumulation predictive capability is critical for refining radioembolization treatment strategies and has the potential to improve therapeutic outcomes [34].

Vascular compliance, characterized as the ability of a blood vessel to stretch without a substantial rise in internal pressure, holds a pivotal role in hemodynamic regulation and substantially influences microsphere accumulation and distribution patterns during radioembolization. The elastic attributes of vascular walls facilitate wave reflection mechanisms, where reflected waves from distal arterial segments can alter local blood pressure and flow dynamics. Consequently, these alterations in hemodynamic conditions exert a significant influence on the microsphere transport and accumulation [50], [51]. Fluid-Structure Interaction (FSI) simulations are employed to model vascular capacitance and its hemodynamic implications within Computational Fluid Dynamics (CFD) frameworks for radioembolization. This FSI study integrates fluid dynamics simulations with the structural mechanics of vessel walls, thereby enabling the accurate representation of phenomena such as arterial compliance and wall deformation [52], [53].

Upon the delivery of a sufficient concentration of microspheres into a blood vessel, their subsequent accumulation can lead to the impedance or complete cessation of blood flow, resulting in stasis. This phenomenon is a critical objective in embolization procedures, particularly when aiming to occlude vessels supplying tumors. The Coupled Computational Fluid Dynamics-Discrete Element Method (CFD-DEM) is a hybrid numerical approach particularly well-suited for simulating microsphere clogging, where cohesive microspheres form obstructions within vessel channels. CFD-DEM method explicitly solves the motion of immersed microspheres within a fluid, while simultaneously accounting for inter-microsphere collisions and fluid-microsphere interactions [54], [55].

CFD solvers are carefully-designed computational tools which numerically solve the governing equations. Despite the high cost of commercial solvers, their ability to handle the complex nature of vasculature, user-friendly interface, high-quality visualization and accurate analysis are highly favorable for radioembolization modeling research. Finite volume solvers (ANSYS Fluent and START-CCM+) discretize the computational domain into control volumes to effectively manage complex flow dynamics, whereas finite element solvers (COMSOL Multiphysics, SimVascular) employ element-based discretization to achieve high precision in modelling patient-specific geometries, catheter tip placement to predict the microsphere distribution. [56] [45].

### 2.1.5 Mesh independence and validation study

A mesh independence study, known as a grid sensitivity study, is the systematic approach to confirm that the numerical solution derived from a CFD simulation remains consistent regardless of the computational mesh resolution. Previous studies on radioembolization have emphasized that sufficient mesh resolution is crucial for accurately capturing the underlying flow physics. Validation is essential in numerical modelling, with validation studies ensuring the correct equations are solved while addressing the limitations of various modelling assumptions. To ensure CFD simulations serve as a dependable research tool, their results must undergo validation by comparing them with real-world data (*in vivo*) or, at minimum, experimental data (*in vitro*). Bomberna et al. [57] performed *in vitro* validation of microsphere distribution in a patient-specific hepatic artery model for CFD studies. Antón et al. [17] validated the CFD results of radioembolization with *in vivo* data.

### 2.1.6 Hemodynamic parameters

The CFD simulation results are used to calculate different hemodynamic parameters described below; their representation and analysis are crucial to derive patient-specific conclusions [58].

Velocity Field: The temporal fluid velocity vector characterizes the flow dynamics within a hepatic artery domain over a defined time between the cardiac cycle. Analyzing local velocity patterns inside the hepatic artery is crucial for predicting areas with elevated microsphere concentrations. [34].

Wall shear stress $\tau_w$: WSS is the tangential force per unit area acting on the vessel wall due to the velocity gradient at the wall surface, which is directly influenced by blood flow dynamics. WSS is a key factor used to analyze vascular remodeling processes, endothelial cell behaviour, and overall vascular integrity [59].

$$\tau_w = \mu \left( \frac{\partial u}{\partial y} \right) \tag{5}$$

where, $\mu$ is dynamic viscosity of blood, and $\left( \frac{\partial u}{\partial y} \right)$ is velocity gradient.

Time averaged wall shear stress (TAWSS): The average WSS value for a single cardiac cycle is called TAWSS. WSS provides the details of vascular behaviour for a specific time frame within a cardiac cycle, but TAWSS is used to observe the WSS changes for the entire cardiac cycle [60].

$$TAWSS = \frac{1}{T} \int_0^T |\tau_w| dt \tag{6}$$

where T is the cardiac cycle time.

### 2.2 AI Approaches for CFD Acceleration

CFD simulations in complex anatomies such as the hepatic or coronary arteries remain computationally expensive due to the need for fine spatiotemporal resolution. To address these challenges, recent advancements in physics-informed neural networks (PINNs) including physics-informed generative adversarial networks (GAN) [61], diffusion models, and transformer-based surrogates [61], [62], [63], [64] have opened new possibilities for accelerating and optimizing CFD workflows. In the context of CFD applications these models integrate physical laws such as the Navier-Stokes equations into the training process of neural networks, allowing them to learn not just from data but also from the underlying physics governing blood flow.

By embedding domain-specific constraints directly into the architecture or loss functions, PINNs can generate high-fidelity velocity and pressure fields with a fraction of the computational cost of traditional solvers. The flexibility of classical PINNs allows them to generalize across varying boundary conditions and anatomical geometries, making them highly effective for surrogate modeling. Extensions of PINNs into generative frameworks further enable fast sampling of diverse flow scenarios, supporting uncertainty quantification and real-time decision support in personalized medicine [65], [66].

### 2.2.1 Classical PINNs: multilayer perceptron-based approaches

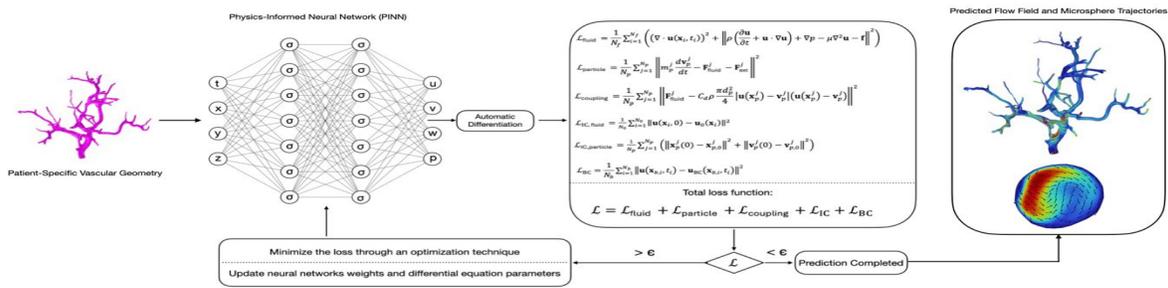

Figure 3: Overview of a Physics-Informed Neural Network (PINN) framework for simulating blood flow and microsphere transport in patient-specific vascular geometries. The PINN takes as input the patient-specific vascular geometry and predicts a flow field and microsphere trajectories that minimize partial differential equations (PDE) residuals at a finite set of collocation points. The loss function includes the fluid, microsphere, initial conditions, and boundary conditions loss terms.

In their basic implementation, PINNs employ a standard multilayer perceptron (MLP) architecture, a feedforward neural network composed of fully connected layers with nonlinear activation functions (Figure 3). This MLP learns mappings from input coordinates, such as spatial and temporal points, to output physical quantities like velocity and pressure. Unlike CFD approaches (Section 2.1), which solve partial differential equations (PDEs) using mesh-based discretization methods (e.g., finite element or finite volume), PINNs treat the problem as a function approximation task. The network is trained to produce outputs that minimize PDE residuals at a finite set of collocation points sampled from a continuous domain, thereby approximating residual minimization across the domain.

What distinguishes PINNs from purely data-driven models is their use of governing physics, in the form of PDEs, as a supervisory signal. This enables them to generalize from sparse clinical data while producing physically consistent 3D flow fields across complex arterial geometries. PINNs leverage automatic differentiation to compute the required spatial and temporal derivatives of the outputs, allowing PDE enforcement without explicit meshing. Mesh-free formulation is particularly advantageous in medical applications, where patient-specific anatomies vary greatly and generating high-quality computational meshes can be difficult [67], [68].

While MLP-based PINNs perform well for deterministic PDEs, they are inherently limited to represent uncertainty, which is central to real-world clinical applications. In radioembolization, patient specific blood flow and microsphere transport involve randomness, ranging from variable inlet velocities and catheter placement to anatomical bifurcations and turbulence. By design MLPs yield deterministic point estimates and cannot capture the distribution of possible outcomes or quantify how uncertainty propagates through the system. As a result, they fail to reflect the probabilistic nature of microsphere delivery and radiation dose deposition. Moreover, MLP-based PINNs require a balanced loss between data and physics terms and training becomes increasingly unstable as the number of stochastic dimensions (random variables or uncertain parameters that represent variability in inputs, boundary conditions, or material properties) increases [69], [70]. This becomes a critical issue in hepatic arterial trees, where geometries are highly individualized and microsphere behavior can be heavily influenced by fine-scale anatomical differences. The lack of inductive bias in MLPs for representing multi-model outputs, combined with their sensitivity to hyperparameters and noise, makes them insufficient for uncertainty-aware clinical modeling.

To address these challenges, recent research has explored combining physical constraints with generative models such as generative adversarial networks (GANs), diffusion models, and attention-based networks, enabling both fast inference and diverse, uncertainty-aware flow predictions [71], [72].

Despite the conceptual promise of these approaches, it is important to note that systematic benchmarks reporting their accuracy and computational performance remain limited. Most current studies demonstrate feasibility on simplified or small-scale vascular geometries, but large-scale, patient-specific validations are still scarce partially due to limited data availability. Reported errors are often context-dependent, for example, early PI-GAN frameworks have shown mean relative errors of 2-5% in velocity fields compared to CFD baselines [49], while diffusion-based surrogates achieve lower residuals and improved stability but have yet to be tested comprehensively on clinical datasets [50], [63]. In many cases, the accuracy of these methods has not been fully quantified beyond proof-of-concept, and when accuracy is unknown, this limitation must be acknowledged explicitly. Going forward, validation should rely on rigorous comparisons against high-fidelity CFD simulations, in vitro flow experiments, and in vivo imaging modalities such as 4D flow MRI or contrast-enhanced CT [76], [77]. Such multi-scale, multi-modal benchmarks will be essential to establish confidence for clinical deployment.

### 2.2.2 Physics-Informed GANs (PI-GANs)

Physics-Informed Generative Adversarial Networks (PI-GANs) [61] offer a powerful extension of classical GANs [73] by embedding known physical laws into the generative modeling framework. While traditional GANs are designed to learn mappings from latent variables to data distributions, PI-GANs incorporate physics-based constraints through additional loss terms, enabling them to generate realistic yet physically consistent outputs. For example, a PI-GAN trained on fluid flow data can generate velocity and pressure fields conditioned on vessel geometry, with discriminator ensuring plausibility and a physics loss penalizing violation of the Navier-Stokes equations (Figure 2).

Unlike deterministic MLP-based PINNs, which produce single-point predictions, PI-GANs are explicitly designed to model distributions of solutions, capturing variability in boundary conditions, anatomy, and flow behavior. The generator outputs a family of plausible solutions driven by latent stochastic variables, allowing PI-GANs to represent uncertainty in microsphere trajectories or blood flow fields across patient-specific hepatic arterial trees. This capability is especially important for $^{90}$Y radioembolization, where treatment planning depends on understanding probabilistic microsphere delivery in highly variable and uncertain vascular conditions. However, although PI-GANs have demonstrated success in generating physically plausible flow patterns, quantitative accuracy across clinically relevant geometries remains under-reported. Current studies often benchmark against synthetic data or simplified flow conditions [49], and systematic error analyses across patient cohorts are still missing. Establishing statistical measures of accuracy (e.g., relative L2 error norms, energy spectra comparisons, wall shear stress distributions) will be critical for assessing their reliability in translational settings.

### 2.2.3 Physics-Informed Diffusion Models

Physics-Informed Diffusion Models (PI-DMs) represent a new class of generative models that combine the expressiveness of diffusion-based architectures [74] with the rigor of physical constraints. Unlike adversarial models that rely on a generator-discriminator setup, diffusion models follow a denoising framework where data are progressively perturbed with noise during training and reconstructed step-by-step during inference. When combined with physical priors, these models can learn to generate physically consistent solutions from incomplete, noisy or low-fidelity inputs.

Trained solely on high-fidelity CFD data, these models learn a generative prior over realistic flow fields. During inference, they can be conditioned on sparse or low-resolution clinical measurements to reconstruct high-resolution velocity and pressure fields that adhere to known physical laws, such as Navier-Stokes equations. This separation between training and conditioning enables PI-DMs to operate even in data-limited clinical settings, while still benefiting from the accuracy and detail of full-order CFD simulations used during training.

Moreover, PI-DMs exhibit greater training stability and coverage of the solution space compared to GAN-based models. GANs are often prone to training instabilities due to adversarial loss, whereas diffusion models are optimized via maximum likelihood, ensuring broader and more consistent representation of the target data distribution [75], [76], [77]. When combined with physics-informed losses applied during the reverse denoising steps, these models can be explicitly guided toward physically valid outputs at each generation step [62], [78]. PI-DMs have not yet been comprehensively benchmarked against gold-standard CFD across full arterial trees. While preliminary reports suggest lower reconstruction error and improved uncertainty quantification compared to GANs [50], most evaluations have been performed on relatively small domains. Future validation strategies should include cross-comparison with CFD-derived hemodynamic biomarkers as well as reproducibility studies across multiple centers and imaging modalities [79]. Without such efforts, the generalizabilty and clinical utility of PI-DMs cannot be fully established.

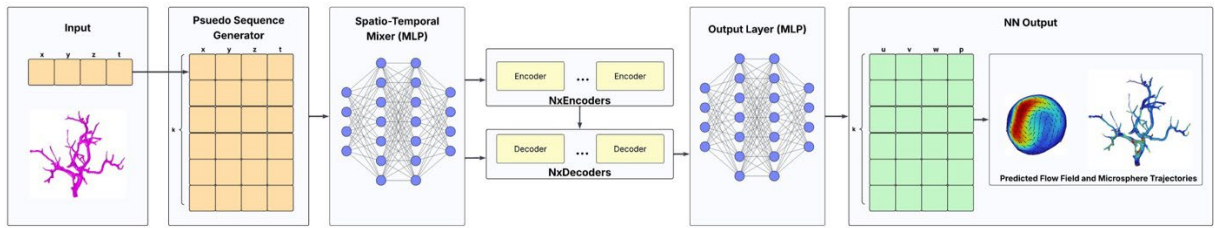

Figure 2: Overview of the adapted PINNsFormer architecture for predicting 3D vascular flow. The model takes spatiotemporal inputs (x, y, z, t), generates pseudo-sequences, and processes them through a spatio-temporal MLP mixer followed by multiple encoder-decoder layers. The output layer predicts velocity and pressure fields (u, v, w, p), enabling reconstruction of the flow field and microsphere trajectories.

2.2.4  Physics-Informed Transformer Models

MLPs, GANs and diffusion models all suffer from limitations when applied to complex, time-dependent systems governed by partial differential equations (PDEs). MLP-based PINNs typically treat each space-time coordinate independently, lacking the ability to model temporal dependencies [80], a critical feature for accurately capturing dynamics in evolving systems such as blood flow. GANs, while capable of generating realistic samples, often suffer from training instability and mode collapse, and typically require low-fidelity input during inference, which may not align with the training distribution [81], [82]. Diffusion models offer more stability and are useful for uncertainty-aware generation, but they rely on iterative denoising steps during inference, which can be computationally expensive and still struggle to capture long-range temporal structure without architectural extensions [83], [84].

Transformer-based architectures [85] offer a compelling alternative by explicitly modeling sequential dependencies through self-attention mechanisms. Originally designed for natural language processing, transformers are inherently suited to learning relationships across time steps, making them ideal for solving parabolic or hyperbolic PDEs where the solution at time $t + \Delta t$ depends on earlier states. In physics-informed settings, transformers can overcome the pointwise limitations of MLPs and the sample inefficiencies of GANs and diffusion models by integrating the physics constraints directly into a sequence-to-sequence modeling framework (Figure 4).

One such architecture, PINNsFormer [64], combines the expressive power of transformers with physics-informed learning. Instead of treating space-time points independently, it generates pseudo-sequences of future time steps for each spatial input and processes them through a Transformer encoder-decoder, thereby explicitly capturing temporal evolution, a critical feature for unsteady flow simulations. The framework also introduces a structured loss decomposition across initial, boundary, and residual points, applying constraints only where physically meaningful. To further enhance temporal modeling, it computes gradients across sequences and incorporates a novel Wavelet activation to represent both periodic and aperiodic behavior. These innovations enable accurate, generalizable, and physically consistent solutions across a wide range of dynamic systems.

Yet, even transformer-based PINNs face unique limitations that hinder their translational impact. While their self-attention mechanisms excel at capturing long-range temporal dependencies, they often struggle with scalability in high-dimensional CFD problems, as the quadratic cost of attention grows prohibitively with spatial resolution. Moreover, without strong physics-specific inductive biases, transformers risk overfitting to training domains and may accumulate errors during long-time rollouts, leading to unstable flow predictions in patient-specific vascular geometries. To establish clinical relevance, validation must extend beyond canonical PDE test cases toward large, standardized datasets and clinically realistic benchmarks. By demonstrating accuracy, stability, and efficiency under such conditions, transformer-based surrogates can evolve into reliable tools for CFD acceleration [86], [87].

### 2.2.5 Physics-Constrained Neural Networks

Physics-constrained neural networks enforce the Navier–Stokes equations as hard constraints, ensuring that predicted velocity and pressure fields strictly satisfy incompressibility, momentum conservation, and other flow physics throughout the domain. Rather than penalizing residuals during training via a loss function, these models encode the PDE structure directly for example, by constructing divergence-free velocity fields using stream functions or solenoidal bases [88], [89], [90]. This architectural embedding guarantees physically admissible solutions by design, improving stability and accuracy especially in fluid flow regimes where enforcement of incompressibility is critical.

## 3 Clinical applications

In the context of radioembolization, translation of digital twins to clinical applications will primarily focus on treatment planning, either before the first intervention where the mapping procedure could be replaced by operating the digital twin, or in between sessions to determine if retreatment is necessary. Several aspects should be considered when discussing translation: how will the digital twin be used? Is it accurate and validated enough for the context of use?

### 3.1 Applications of CFD to radioembolization treatment planning

As pointed out in previous sections, CFD has been extensively applied to radioembolization treatment planning, studying the effect of catheter positioning, angle, distance from arterial bifurcations, and other features[15], [16], [17], [18], [19], [14]. While CFD has not been translated clinically to inform treatment yet, its application would consist in sampling injection locations, injection rate, and $^{90}$Y activity to identify the best combination to obtain the desired dose distribution in the liver (healthy tissues and tumor regions), as shown in Figure 5. Initial studies already improved the understanding of which factors most impact the microsphere distribution (and subsequent the dose distribution) and should be carefully considered during treatment (catheter position with the arterial tree, angle, and injection timing with respect to the cardiac cycle).

## 3.2 Applications of PINNs to CFD-based digital twins

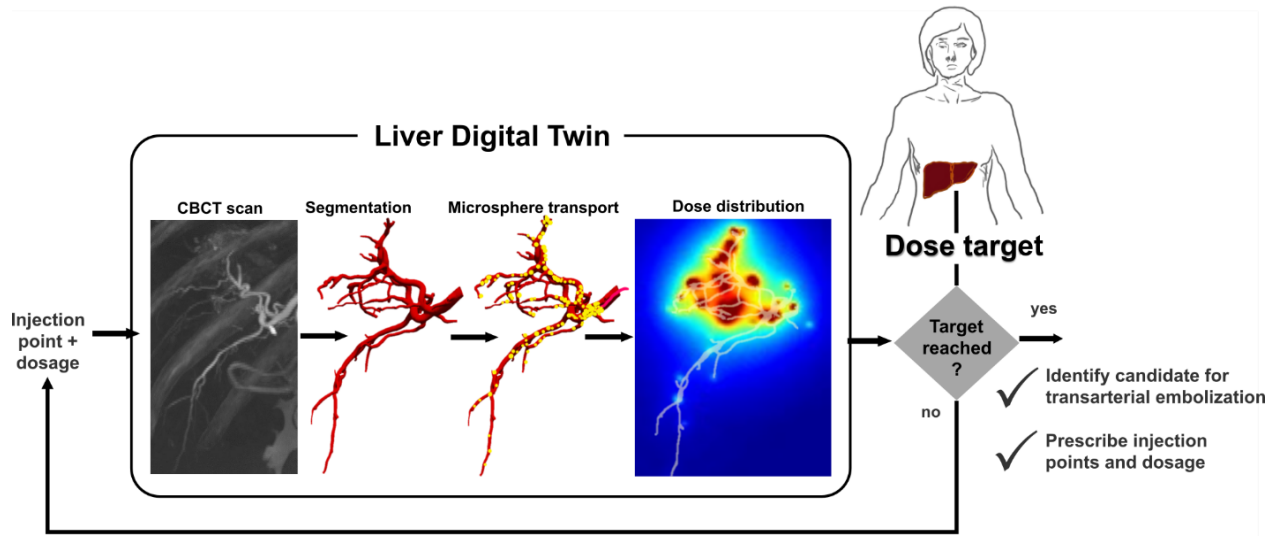

**Figure 3.** Example of liver digital twin for radioembolization treatment planning based on CFD to model $^{90}$Y microsphere transport in the liver. The digital twin predicts the dose distribution for a set of conditions (inject point + dosage), which is compared to the target defined by the physician. Conditions are updated until an acceptable dose output is obtained and ultimately define a treatment plan.

One of the primary goals of the digital twin for radioembolization is to optimize treatment, which translates into maximizing the dose to the tumor(s) while minimizing the dose to the rest of the liver (specific segments, lobe, or whole liver). The target doses to these regions should be defined by the interventional radiologist performing radioembolization, based on well-established criteria for radioembolization to avoid lung shunting and radiation-induced liver disease. To progress from a CFD model or PINN to a full liver digital twin, personalization (e.g. with data from imaging) and incorporation of the twin into an algorithm that can be updated to estimate output variables is necessary. An example of *in silico* treatment planning for radioembolization is shown in Figure 5. In this proposed theranostic digital twin, the optimization task consists of sampling combinations of injection points and dosages and simulate the corresponding dose distribution with the liver digital twin. Each injection will distribute microspheres within the arterial tree, contributing to the total radiation dose distribution. The output of the digital twin is the total dose distribution, which is evaluated for various injection points and activities and compared against the dose target until the difference is acceptable or until the patient is considered not eligible (maximum dose to healthy liver exceeded). Figure 5 illustrates the liver digital twin built on a CFD model, however this model can easily be replaced by a PINN, PI-GAN or other generative AI-based model that bears the same prediction capability.

## 3.3 Accuracy and validation

Once the application and the context of use of the digital twin have been defined, it is important to characterize the accuracy and reliability of the model and determine what validation steps should be taken. Similar to other computational and experimental models, digital twins are only valid over a given range of environment and system parameters and can be considered accurate enough when they produce a reliable output within this range. If CFD is used as the mechanism to simulate the microsphere distribution, the first validation stage consists of comparing the results with benchmark CFD studies.

The accuracy of the dose distribution will depend on many factors, which include the spatial resolution of the images used to extract the 3D vascular mesh (~ 200 µm), CFD assumptions (e.g. microsphere interactions, spatio-temporal considerations such as injection rate, cardiac cycle etc), and accounts for the particle physics modeling uncertainty. In radioembolization, the accuracy is hard to quantify due to the lack of high-precision ground truth, limited by the resolution of the $^{90}$Y PET images or $^{99m}$Tc-MAA SPECT (3 mm, and over 5 mm respectively), registration between structures and regions of interest segmentation.

Validation is not a binary process but rather a continuum which can be implemented using a verification, validation, and uncertainty quantification framework (VVUQ) [91]. A systemic digital twin can be validated and optimized at the organ level then at a multiorgan level including cross-interactions (e.g. cross dose between organs in dosimetry applications). For our liver radioembolization digital twin , metrics such as going from flow fields computed by a CFD simulation or PINN, microsphere distribution, $^{90}$Y activity distribution or dose distribution in liver segments could be compared with measured data. Voxelwise comparisons might be considered, although extremely challenging due to the limitations discussed in the previous paragraph.

*3.4  Translation*

The goal of the theranostic liver digital twin is to find the best combinations of injection locations within the liver arterial tree and the activity injected at each location satisfying the physician constraints in terms of dose to the tumor, liver segments, healthy liver. Those different regions are called target (tumor) which should receive the highest dose and the organs at risk (OAR), which dose should be limited to avoid toxicity and side effects such as decreased liver function, radiation-induced liver disease [9].

The digital twin's output, as described in Section 3, is a predicted 3D dose distribution in the liver. In the future, this dose distribution could expand to the predicted dose to the lungs through arterio-venous shunting modeling for given injection locations and activity. Modeling several injection points can be done sequentially as injections would be done in the clinics and do not require modeling the interaction between microspheres from different injections, simplifying the process. Considering real-time updates of the digital twin, modeling the effect of a given injection on the pressure and flow conditions (e.g. tumor clustering) could allow more precise prediction of the cumulative dose distribution when several injections are simulated. The capacity to integrate updated data dynamically should be embedded in the digital twin mode design from its inception, to allow efficient data flow. Similarly, a range of activities can be simulated from a unique injection location simulation, assuming the number of microspheres stays low enough to not change the flow pattern and the microsphere distribution between vessel branches. The sensitivity of a particular geometry (for a given patient) to the volume of microspheres will likely vary between patients depending on the vessel size and the type of microsphere.

After performing initial treatment planning, the digital twin can be also used in conjunction with post-treatment imaging to decide if sufficient dose was delivered to the tumor(s) and if toxicity limits were not exceeded. Additional injections may be planned, either with a updated digital twin to account for hemodynamics changes, tumor volume and vascular changes, liver vascular reorganization.

The power of digital twins and computational modeling is their multiphysics nature, which makes it possible to include biophysics and biologically-informed models that could predict tumor and liver response to treatment, including toxicity, provided that the digital twin incorporates enough biomarkers to enable this predictive capability. Other clinical applications in the context of radioembolization could be considered, such as brain tumor radioembolization where guidance to avoid eloquent white matter is critical [92].

# 4 Conclusions and future perspectives

Radioembolization is an ideal application for digital twins, as modeling an organ is simpler than an entire system. Two main challenges remain in translating radioembolization digital twins into clinical practice: personalizing the model with patient-specific data (e.g. hemodynamics for boundary conditions) and accelerating computation to produce treatment plans within a few days. The small number of parameters to optimize and the underlying physics of radioembolization (CFD) make PINNs promising to accelerate computation, as demonstrated in cardiovascular applications. Future steps include visualization, augmented reality for real-time treatment planning, and integration in clinical software. Interventional radiology extensively uses 3D visualization software, making this integration within reach.

# Acknowledgments

We acknowledge funding from NCI ITCR U01CA289068 grant.